%% file: root.tex
\documentclass[letterpaper, 10 pt, conference]{cls/ieeeconf}

\IEEEoverridecommandlockouts
\let\labelindent\relax

\usepackage[normalem]{ulem}	                        
\usepackage[table,usenames,dvipsnames]{xcolor}      
\usepackage{enumitem}                               
\usepackage{extarrows}                              
\usepackage[noadjust]{cite}                         

\usepackage{amsmath,amssymb,amsfonts,amsthm, dsfont} 		
\usepackage{algorithm,algorithmicx,listings}    
\usepackage[noend]{algpseudocode}			          
\usepackage{mathtools} 

\usepackage{graphicx,tabularx,subcaption}
\usepackage[export]{adjustbox}

\usepackage[font={small}]{caption}
\usepackage[titletoc,title]{appendix}

\usepackage[breaklinks=true, colorlinks, bookmarks=true, citecolor=Black, urlcolor=Violet,linkcolor=Black]{hyperref}

\theoremstyle{definition}
\newtheorem{definition}{Definition}

\newtheorem*{assumption*}{Assumption}
\newtheorem*{problem*}{Problem}

\theoremstyle{remark}
\newtheorem{remark}{Remark}
\newtheorem*{solution*}{Solution}

\def\thetitle{Comparison between safety methods control barrier function vs. reachability analysis}
\def\theauthor{Zhichao Li}
\def\thekeywords{safety control, control barrier function, reachability analysis}

\hypersetup{
  pdfauthor={\theauthor},%
  pdftitle={\thetitle},%
  pdfsubject={\thetitle},%
  pdfkeywords={\thekeywords}
}

\IEEEoverridecommandlockouts

\title{\LARGE \bf \thetitle}

\author{%
	Zhichao Li, 
	\\
	Department of Electrical and Computer Engineering, University of California, San Diego\\
	zhichaoli@ucsd.edu
}
\input{cls/sym.tex}

\begin{document}
\maketitle
\begin{abstract}
This report aims to compare two safety methods: control barrier function and Hamilton-Jacobi reachability analysis. We will consider the difference with a focus on the following aspects: generality of system dynamics, difficulty of construction and computation cost. 
A standard Dubins car model will be evaluated numerically to make the comparison more concrete. 
\end{abstract}

\input{tex/Introduction.tex}

\input{tex/Overview.tex}
\input{tex/Problem.tex}

\input{tex/Background.tex}
\input{tex/Connection.tex}
\input{tex/Comparison.tex}

\input{tex/CaseStudy.tex}
\input{tex/Conclusion.tex}

{\small
\bibliographystyle{cls/IEEEtran}
\bibliography{bib/ref.bib}
}
\end{document}

%% file: cls/sym.tex
\newcommand{\calC}{{\mathcal C}}
\newcommand{\calD}{{\mathcal D}}

\newcommand{\calF}{{\mathcal F}}
\newcommand{\calG}{{\mathcal G}}

\newcommand{\calK}{{\mathcal K}}
\newcommand{\calL}{{\mathcal L}}

\newcommand{\calR}{{\mathcal R}}

\newcommand{\calU}{{\mathcal U}}

\newcommand{\calX}{{\mathcal X}}

\newcommand{\bfb}{\mathbf{b}}

\newcommand{\bfd}{\mathbf{d}}

\newcommand{\bff}{\mathbf{f}}
\newcommand{\bfg}{\mathbf{g}}

\newcommand{\bfk}{\mathbf{k}}

\newcommand{\bfp}{\mathbf{p}}

\newcommand{\bfu}{\mathbf{u}}

\newcommand{\bfx}{\mathbf{x}}

\newcommand{\bfxi}{\boldsymbol{\xi}}

\newcommand{\bfA}{\mathbf{A}}

\newcommand{\bfG}{\mathbf{G}}
\newcommand{\bfH}{\mathbf{H}}

\newcommand{\bbR}{\mathbb{R}}

\newcommand*{\Bzero}{\mathbf{0}} 

\newcommand{\scaleMathLine}[2][1]{\resizebox{#1\linewidth}{!}{$\displaystyle{#2}$}}
\newcommand{\prl}[1]{\left(#1\right)}
\newcommand{\brl}[1]{\left[#1\right]}
\newcommand{\crl}[1]{\left\{#1\right\}}

\DeclareMathOperator*{\argmax}{argmax}   
\DeclareMathOperator*{\argmin}{argmin}   

\DeclarePairedDelimiterX{\norm}[1]{\lVert}{\rVert}{#1}

\newcommand*{\nRbtState}{n}
\newcommand*{\nInput}{m}

%% file: tex/Introduction.tex
\section{Introduction} 
\label{sec:introduction} 
Advances in embedded sensing and computation have enabled robot applications in all kinds of environments and in close interaction with humans, including autonomous transportation, inspection and cleaning services, and medical robotics. Designing provably safe controllers becomes key challenge to these safety-critical applications. Moreover, many of these systems are affected by disturbance in real applications. How to formally verify safety of designed autonomous system becomes active research area in recent years. Researchers have been developed many tools to handle this problem, such as model predictive control (MPC), reference governor, reachability analysis, barrier function. Although lots of literature exists for each individual approach, the comparison between them is less explored. In this report, we are aiming to compare the last two approaches with focus on Hamilton-Jacobi (HJ) reachability analysis and Control Barrier Function (CBF) method.


Hamilton-Jacobi (HJ) reachability analysis is a formal verification method for safety of dynamical systems. It computes the reach-avoid set which is the set of states from which the dynamical system can be driven to a target set while avoid unsafe set subject to bounded uncertainty. With modern numerical tools \cite{mitchell2002_thesis, mitchell2007toolbox}, one can computes different types of reachable sets. These tools have been successfully applied in various problems, for example, differential games, safe motion planning and optimal control problems \cite{HJ_tutorial}. Reachable set computations often involve solving an HJ partial differential equation (PDE) on grid of the state space (system states, control, and disturbance), resulting in an exponential scaling of computational complexity with respect to system dimensionality.  This makes HJ reachability analysis becomes computationally intractable as the state space dimension increases.  However, by exploiting system structures, system decomposition \cite{HJA_decomposition_chen2017} technique has significantly mitigated this problem and apply HJ analysis to 10-D quadrotor model \cite{herbert2017fastrack}. 

Control barrier functions (CBFs) are level-set functions used to provides formal safety guarantees for nonlinear control affine systems. New formulation of CBF provides a less conservative and flexible constraint enforcement by allowing barrier function to grow when is far away from the boundary of safe set \cite{CBF_ames2014control,CBF_ames2017TAC}. Recent formulation \cite{CBF_ames2014control} combing control Lyapunov function (CLF) and CBF in an optimization via a quadratic program (QP) can jointly consider safety constraints via CBFs and system stability requirements through CLFs. This QP based controller can be solved online efficiently and has been successfully used in various multi-input-multi-output safe-critical systems like lane-keeping for autonomous car, Segway balancing, bipedal walking robots, quadrotors, multi-robot formulation \cite{CBF_ames2014control,CBF_Ames2015_swarm,CBF_ames2014rapidly,CBF_quadrotor,CBF_ECBF_nguyen2016ACC}. Despite the great success, developing a CBF is challenging for general systems with high-relative degree systems \cite{CBF_ames2019ECC}. This problem becomes even challenging when the system dynamics is subject to unknown disturbance.

They are good literature review paper for both methods. For example, Bansal et al. \cite{HJ_tutorial} gives formal and self-contained introduction to reachability theory as well as recent developments that help overcome the curse of dimensionality; Ames et al. \cite{CBF_ames2019ECC} is a comprehensive survey paper about control barrier function-based methods. Main technical results are given with various applications. However, the difference and limitations of these two methods is not addresses in detail. The goal of this report is two-fold. First, we aim to show both strengths and weakness of both methods so practitioner can pick the suitable tool for their applications. Second, we hope this report can inspire future researcher to find interesting directions.

%% file: tex/Overview.tex
\section{Overview of the Paper}
In this report, we will consider two safety methods: control barrier function (CBF) and Hamilton-Jacobi reachability analysis (HJ-RA). This report will consist of following parts:
\begin{itemize}
	\item background of each method
	\item connection of the two methods
	\item comparison of these methods in terms of generality of system dynamics, difficulty of construction, computation cost. 
	\item challenges and practical issue arises from two methods
	\item case study of Dubins car
	\item conclusion and summary for both approaches
\end{itemize}

%% file: tex/Problem.tex
\section{Problem Formulation}
\label{sec:problem_formulation}
Consider a general nonlinear dynamical system operating in an environment subject to external disturbance satisfying the following dynamics:
\begin{equation}
\label{eq:sys_pf}
\begin{aligned}
\dot{\bfx} &= \bff(\bfx, \bfu, \bfd)
\end{aligned}
\end{equation}
where $\bfx \in \calX \subset \bbR^\nRbtState$ is the state, $\bfu \in \calU \subset \bbR^\nInput$ is the admissible control input, and $\bfd \in \calD \subset \bbR^\nRbtState$ is the disturbance might be an exogenous noise or model uncertainty. We assume $\bff$ are locally Lipschitz, so that there exists a unique trajectory of \eqref{eq:sys_pf}. Let $\bfxi_{\bfx_0 , t_0}^{\bfu, \bfd}(t)$
denotes the system trajectory at time $t$ by starting at time $t_0$ with initial state $\bfx_0$ and applying input functions $\bfu(\cdot)$ and $\bfd(\cdot)$ over time horizon $\brl{t_0, T}$.  The environment contains a set $\calG$ that is the goal configuration that the agent would like to reach and unsafe region $\calF$ which would lead to harmful behaviors.
We would like to find the control inputs $\bfu(\cdot)$ that can drive the system reach the goal region $\calG$,  despite the worst-case disturbance and while avoiding unsafe set $\calF$ at all times.

%% file: tex/Background.tex
\section{Background}
\label{sec:background}
\subsection{Hamilton Jacobi reachability analysis}
Hamilton-Jacobi reachability analysis (HJ-RA) is a formal verification method for safety of dynamical systems. The key concept of this approach is backward reachable tube and backward reach-avoid tube.
\begin{definition} \label{def:BRT}
	\textbf{Backward Reachable Tube (BRT)}. BRT is the set of initial states for which the agent acting optimally and under worst-case disturbances, will eventually reach the target set $\calL$ within the time horizon $\brl{t_0, T}$\footnote{Here, terminal time $T=0$ since we are computing value function and reachable set backward in time.}:
	\begin{equation} \label{eq:brt}
	\scaleMathLine{
	\calR^{BRT}_{t_0}(\calL) = \crl{\bfx_0 \in \calX: \exists \bfd(\cdot), \forall \bfu(\cdot), \exists t \in \brl{t_0, T}, \bfxi_{\bfx_0 , t_0}^{\bfu, \bfd}(t) \in \calL}
	}
	\end{equation}
\end{definition}
\begin{remark}
In above definition, non-anticipative strategy for disturbance is assumed, more detail can be found in \cite{HJ_time_varying}.
\end{remark}
BRT contains all possible unsafe initial conditions that potentially lead to failure. In some literature, BRT has been called robust inevitable backward reachable tube.
%
In HJ-RA, the computation of BRT is formulated as a zero-sum differential game between control and disturbance which can be solved using principle of dynamic programming. First, define the unsafe set $\calF$ as sub-zero level function of some function $l(\bfx)$, i.e, $\calF = \crl{\bfx: l(\bfx) \leq 0}$ and treat it as target set $\calL$. Typically, the signed distance function to $\calF$ is used \cite{HJ_tutorial}.  Then, we can define reward function as the pointwise in time minimum distance of trajectory to $\calF$ over time:
\begin{equation} \label{eq:HJ_cost}
J(\bfx_0, t_0, \bfu(\cdot), \bfd(\cdot)) = \min_{t \in \brl{t_0, T}} l(\bfxi^{\bfu, \bfd}_{\bfx_0, t_0}(t)).
\end{equation}
The associated value function is defined as:
\begin{equation}
V_{BRT}(\bfx, t) = \inf_{\bfd(\cdot)} \sup_{\bfu(\cdot)} \crl{J\prl{\bfx, t, \bfu(\cdot), \bfd(\cdot)}}
\end{equation}
This value function can be computed using dynamic programming, which requires to solve the following Hamilton-Jacobi-Isaacs (HJI) equations:
\begin{equation} \label{eq:HJI_BRT}
\begin{split}
\min \big\{ D_t V_{BRT}(\bfx, t) +   H(\bfx, t) &, l(\bfx) - V_{BRT}(\bfx, t) \big\} = 0, \\
V_{BRT}(\bfx, T) &= l(\bfx) 
\end{split}
\end{equation}
where 
$H(\bfx, t) = \inf_{\bfd} \sup_{\bfu} D_\bfx V_{BRT}(\bfx, t) \cdot \bff(\bfx, \bfu, \bfd)$ is the Hamiltonian. 
Once the value function is computed, the BRT is given as the sub-zero level set of the value function, i.e., 
\begin{equation}
\calR^{BRT}_{t_0}(\calF) = \crl{\bfx: V_{BRT}(\bfx, t_0) \leq 0}.
\end{equation}
The corresponding optimal safety control can be derived as
\begin{equation} \label{eq:HJ_optimal_BRT}
\bfu^*_{BRT}(\bfx, t) = \argmax_{\bfu} \min_{\bfd} D_\bfx V_{BRT}(\bfx, t) \cdot \bff(\bfx, \bfu, \bfd)
\end{equation}
Suppose we are given a reference control $\bfu_{ref}(t, \bfx)$ for goal reaching which is not necessarily safe with respect to $\calF$ subject to disturbance. Using optimal control law~\eqref{eq:safe_ctrl_brt}, we can use the following \textbf{BRT-based least restrictive control} policy:
\begin{equation} \label{eq:safe_ctrl_brt}
\pi_{BRT}\prl{t, \bfx} =\begin{cases}
\bfu_{ref}(t, \bfx)  &\text{if} \; V_{BRT}(\bfx, t) \geq \epsilon_{BRT}, \\
\bfu^*_{BRT}(\bfx, t),  &\text{otherwise}.
\end{cases}
\end{equation} 
where $\epsilon_{BRT} > 0$ is some small number to prevent numerical errors. As we already seen, Hamilton Jacobi reachability analysis can also be used to quantify unsafe initial states, it can also be formulated to solve a reach-avoid problem. 
\begin{definition} \label{def:BRAT}
	\textbf{Backward Reach-Avoid Tube (BRAT)}. BRAT is the set of initial states from which there exist a control consequence that can drive the agent towards goal set $\calG$ despite worst-case disturbances, without ever entering unsafe set $\calF$. Mathematically,
	\begin{equation} \label{eq:brat}
	\begin{split}
			\calR^{BRAT}_{t_0}(\calG, \calF) &= \{\bfx_0 \in \calX: \exists \bfu(\cdot), \forall \bfd(\cdot), \\
			& \exists s \in \brl{t_0, T}, \bfxi^{\bfu, \bfd}_{\bfx_0, t_0}(t) \in \calG, \\
		 	&\forall t \in \brl{t_0, s}, \bfxi^{\bfu, \bfd}_{\bfx_0, t_0}(t) \notin \calF \}
	\end{split}
	\end{equation}
	where, target set $\calL \equiv \calG = \crl{\bfx: l(\bfx) \leq 0}$ and unsafe set $\calF = \crl{\bfx: g(\bfx) \geq 0}$.
\end{definition}

The HJI Variational Inequality (HJI VI) for BRAT is defined as below: 
\begin{multline} \label{eq:HJI_BRAT}
0 	= \max \bigg\{\min \big\{ l(\bfx) - \tilde{V}(\bfx, t), D_t \tilde{V}(\bfx, t) + \tilde{H}	\big\}   \\
  g(x) - \tilde{V}(\bfx, t)  \bigg\} \\
\tilde{V}(\bfx, T) = \max \crl{l(\bfx), g(\bfx)}
\end{multline}
where the associated value function and Hamilton are
\begin{multline}
\tilde{V}(\bfx, t) = \inf_{\bfu} \sup_{\bfd} \min_{t \in \brl{t_0, T}} \max \bigg\{l\big(\bfxi^{\bfu, \bfd}_{\bfx, t_0}(t)), \\\max_{s \in \brl{t_0, t}} g\big(\bfxi^{\bfu, \bfd}_{\bfx, t_0}(s) \big) \bigg\} \\
\tilde{H}(\bfx, t) = \inf_{\bfu} \sup_{\bfd} D_\bfx \tilde{V}(\bfx, t) \cdot \bff(\bfx, \bfu, \bfd)
\end{multline}
The optimal control law is:
\begin{equation} \label{eq:HJ_optimal_BRAT}
\bfu^*_{BRAT}(\bfx, t) = \argmin_{\bfu} \sup_{\bfd} D_\bfx \tilde{V}(\bfx, t) \cdot \bff(\bfx, \bfu, \bfd)
\end{equation}

In addition, we can define \textbf{BRAT-based least restrictive controller} as follows:
\begin{equation} \label{eq:safe_ctrl_brat}
\pi_{BRAT}\prl{t, \bfx} =\begin{cases}
\bfu_{ref}(t, \bfx)  &\text{if} \; \tilde{V}(\bfx, t) \geq \epsilon_{BRAT}, \\
\bfu^*_{BRAT}(\bfx, t),  &\text{otherwise}.
\end{cases}
\end{equation} 

\subsection{Control Barrier Function}
\label{sec:background_cbf}
For control barrier function, we restrict us to deterministic control-affine system which leads to efficient Quadratic Program (QP) formulation that can be efficiently solved online. We will discuss the generalization to stochastic system in later section.
Consider a control-affine nonlinear dynamical system,
\begin{equation}
\label{eq:sys_ctrl_affine}
\begin{aligned}
\dot{\bfx} &= \bff(\bfx) + \bfG(\bfx) \bfu,
\end{aligned}
\end{equation}
where $\bfx \in \calX$  is the state, $\bfu \in \calU$ is the control input, and $\bff(\bfx)$\footnote{Here, symbol $\bff$ is overloaded for simplicity. It should be easy to tell the real meaning from context.}, $\bfG(\bfx)$ are locally Lipschitz continuous functions. Let us first introduce control Lyapunov function (CLF) which inspires the modern formulation of control barrier function (CBF). Then we will describe the popular CLF-CBF QP optimization-based controller which can balance safety and stability goal at the same time enabling lots of recent applications~\cite{CBF_ames2019ECC}.
\begin{definition} \label{def:CLF}
	A function $V(\bfx): \calX \mapsto \bbR_{\geq 0}$ is a \emph{control Lyapunov function} for system~\eqref{eq:sys_pf} if it is positive definite, $V(\bfx) > 0, \forall \bfx \in \calX \setminus \crl{\Bzero}$, $V(\Bzero) = \Bzero$, and satisfies:
	\begin{equation} \label{eq:CLF_IE}
	\inf_{\bfu \in \calU(\bfx)} \brl{\calL_{\bff}V(\bfx) + \calL_{\bfG}V(\bfx) \bfu} 
	\leq -\gamma \prl{V(\bfx)},
	\end{equation}
	where $\calL_\bff V(\bfx)$ is the Lie derivative of $V(\bfx)$ along $\bff(\bfx)$, $\calL_\bfG  V(\bfx) \bfu \coloneqq \sum_{j=1}^{\nInput} u_j \calL_{\bfg_j} V(\bfx)$ with $\bfg_j(\bfx)$ being the $j$-th column of $\bfG(\bfx)$, and $\gamma$ is a class-$\calK$ function.
\end{definition}
Def.~\ref{def:CLF} allows us to characterize the set of stabilizing controllers associated with $V(\bfx)$ at any $\bfx \in \calX$:
\begin{equation} \label{eq:Kclf}
\scaleMathLine{\calK_{clf}(\bfx) \coloneqq \crl{\bfu \in \calU(\bfx) \mid \calL_{\bff}V(\bfx) 
		+ \calL_{\bfG}V(\bfx) \bfu \leq -\gamma \prl{V(\bfx)}}}.
\end{equation}
If there exists a CLF $V(\bfx)$, then any Lipschitz continuous feedback controller $\bfk(\bfx) \in \calK_{clf}(\bfx)$ asymptotically stabilizes the system~\eqref{eq:sys_ctrl_affine} to origin~\cite[Thm.~1]{CBF_ames2019ECC}.
In most the literature, ensuring safety using CBF refers to guaranteeing forward invariance of a safe set. 
Ames et al.~\cite{CBF_ames2017TAC} propose the concept of a zeroing control barrier function that allows characterizing the controllers that render a set $\calC$ forward invariant with respect to \eqref{eq:sys_ctrl_affine}.
We will define them formally at below.

\begin{definition}
	A set $\calC \subset \calX$ is \emph{forward invariant} with respect to dynamics~\eqref{eq:sys_ctrl_affine} if for every initial condition $\bfx \in \calC$, the system trajectory satisfies $\bfxi_{\bfx_0 , t_0}^{\bfu}(t) \in \calC$ for all $t \geq t_0$. 
\end{definition}

\begin{definition} \label{def:CBF}
	Let $\calC \coloneqq \crl{\bfx \in \calX \mid b(\bfx) \geq 0}$ be the superlevel set of a function $b \in C^1(\calX,\bbR)$. Then, $b$ is a \emph{zeroing control barrier function} (ZCBF) for $\calC$ if there exist an extended class $\calK_\infty$ function $\alpha$ such that for system~\eqref{eq:sys_ctrl_affine}:
	\begin{equation} \label{eq:CBF_IE}
	\sup_{\bfu \in \calU} \brl{\calL_{\bff} b(\bfx) + \calL_{\bfG} b(\bfx)\bfu} \geq -\alpha \prl{b(\bfx)}, \quad \forall \bfx \in \calC.
	\end{equation}
\end{definition}

Def.~\ref{def:CBF} allows us to characterize the set of all safe controllers at any $\bfx \in \calX$:
\begin{equation} \label{eq:Kcbf}
\calK_{cbf}(\bfx) \coloneqq \crl{\bfu \in \calU(\bfx) \mid \calL_{\bff} b(\bfx)
	+ \calL_{\bfG} b(\bfx)\bfu \geq -\alpha \prl{b(\bfx)}}.
\end{equation}
If $b$ is a ZCBF for $\calC$ and $\nabla b(\bfx) \neq 0$ for all $\bfx \in \partial \calC$, then any Lipschitz continuous controller $\bfk(\bfx) \in \calK_{cbf}(\bfx)$ for~\eqref{eq:sys_ctrl_affine} renders $\calC$ safe~\cite[Thm.~2]{CBF_ames2019ECC}. Moreover, if the safe set $\calC$ is compact, $\calC$ is forward invariant if and only if it admits a ZCBF~\cite{CBF_ames2017TAC}.

\textbf{Optimization-based Stabilizing Safe Control.}
\label{sec:CLF-CBF-QP}
Note that the sets of stabilizing controllers $\calK_{clf}(\bfx)$ and safe controllers $\calK_{cbf}(\bfx)$ are defined by affine constraints in $\bfu$. This observation allows the formulation of control synthesis as a quadratic program in which stability and safety properties are captured by the affine CLF and CBF constraints, respectively. This formulation is known as a CLF-CBF QP \cite{CBF_ames2017TAC}:
\begin{align} \label{eq:CLF_CBF_QP}
\bfk^*(\bfx) 	= &\argmin_{\prl{\bfu, \delta} \in \bbR^{m+1}} \bfu^{\top} \bfH(\bfx) \bfu + p \delta^2 \\
&\quad\text{s.t.}\quad \calL_{\bff}V(\bfx) + \calL_{\bfG}V(\bfx) \bfu \leq -\gamma  \prl{V(\bfx)} + \delta 	\notag\\
&\quad\phantom{\text{s.t.}}\quad \calL_{\bff}b(\bfx) + \calL_{\bfG}b(\bfx)\bfu \geq -\alpha \prl{b(\bfx)} 	\notag\\
&\quad\phantom{\text{s.t.}}\quad \bfA_\bfu (\bfx) \bfu \leq \bfb_\bfu (\bfx),\notag
\end{align}
where $\bfH\prl{\bfx}$ is any positive definite matrix pointwise in $\bfx$ penalizing control effort and $\delta$ is a slack variable ensuring the QP feasibility\footnote{Assuming CBF inequality is always satisfiable.} by giving preference to safety over liveness, controlled by the scaling factor $p > 0$. The admissible control set is a polytope $\underline{\calU}(\bfx) := \crl{ \bfu \in \mathbb{R}^m \mid \bfA_\bfu (\bfx) \bfu \leq \bfb_\bfu (\bfx)}$ which can encode linear state constraints easily.

\textbf{Minimally invasive CLF-QP controller.}
In many scenarios, a well-performing stabilizing control law $\bfk(\bfx)$ is already available, e.g., obtained via MPC, back-stepping, machine learning, or other techniques, but needs to be modified online to ensure safety (it may be the case that $\bfk(\bfx) \notin \calK_{cbf}(\bfx)$ for some $\bfx \in \calC$). In this case, one does not need to enforce the CLF constraint and may modify the controller $\bfk(\bfx)$ minimally using a CBF QP to guarantee safety:
\begin{align} \label{eq:MIX_CBF_QP}
\bfk^*(\bfx) 	= &\argmin_{\bfu \in \bbR^m} & & \norm {\bfu - \bfk(\bfx)}_2^2 \\
&\text{subject to}  &  &\calL_{\bff}b(\bfx) + \calL_{\bfG}b(\bfx)\bfu \geq -\alpha \prl{b(\bfx)} 	\notag\\
& & &\bfA_\bfu (\bfx) \bfu \leq \bfb_\bfu (\bfx). \notag
\end{align}

%% file: tex/Connection.tex
\section{Connection}
\begin{figure}[t]
	\includegraphics[width=\linewidth]{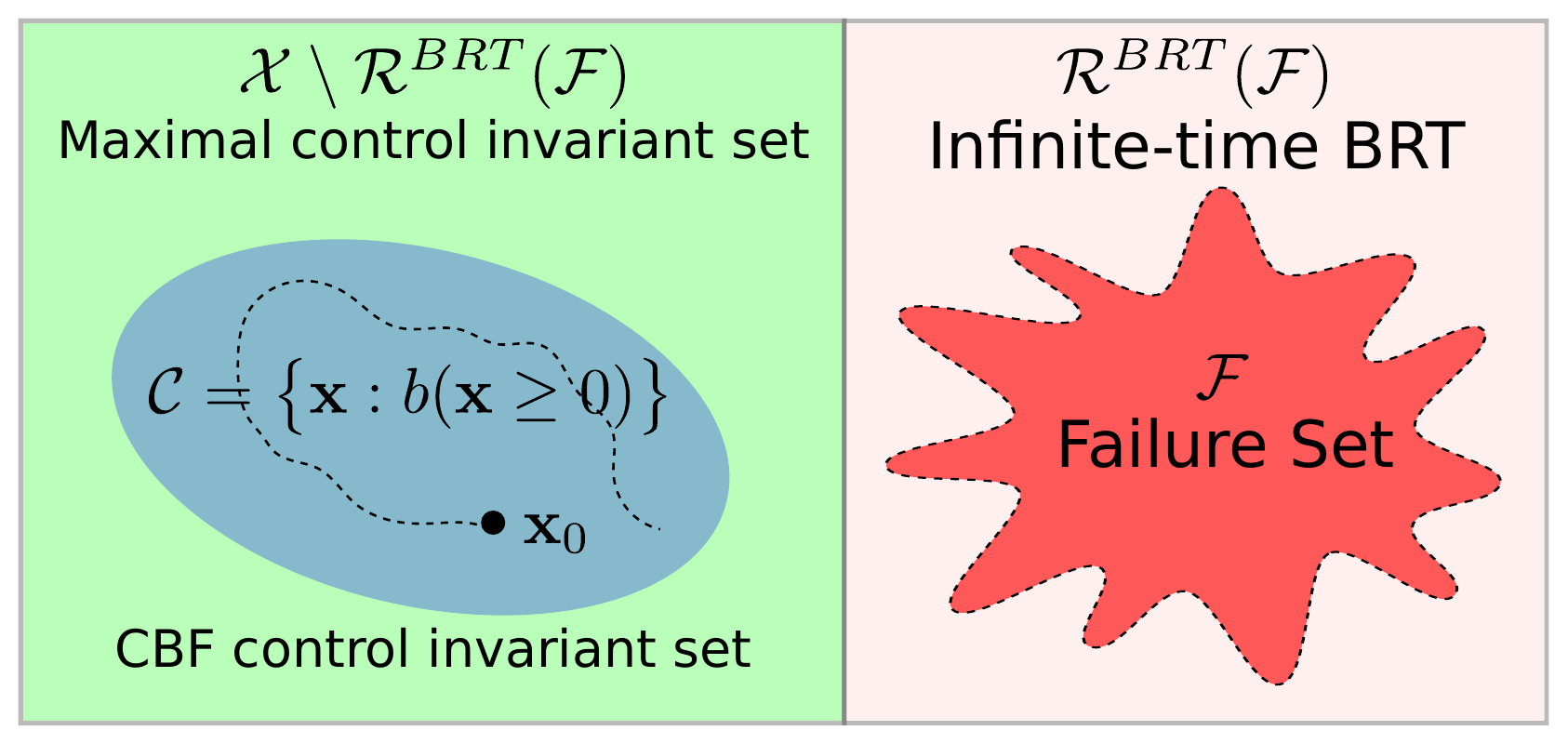}
	\caption{Geometric relationship of safe sets characterized by CBF (blue) and HJ-RA (Green) for deterministic control-affine systems.}
	\label{fig:geometric_connection}
\end{figure}
In this section, we will discuss connections between CBF and HJ analysis and in the following three aspects. 
First, how safe sets defined by two methods related? Second, can we build a CBF from HJ reachability analysis?
Finally, how stability can be ensured in terms of each method.
Since most of CBF literature is focus on deterministic control-affine system dynamics, our discussion in this section is restricted to this as well. Let us start with the characterization of safe set of both methods. In most of applications, the scale of the disturbance scale is smaller than control. Therefore, we assume that as initial time $t_0 \rightarrow -\infty$ \footnote{Here, we consider terminal time $T=0$} BRT will converge. The dependency on $t_0$ of value function BRT will be dropped in this case.

\textbf{Safe set characterization.} Let us start with the first question. If we take the complement of this infinite-time BRT, we get the maximal robust control invariant set (also named viable set \cite{CBF_Segway}) denoted as $\calX \setminus \calR^{BRT}(\calF)$. This set contains all initial conditions that is safe, i.e, if system trajectory start within it, there exist a control policy can drive the system away from the failure set $\calF$ no matter how disturbance against us. 
From Definition~\ref{def:CBF}, we know that, safe set $\calC$ is defined as superlevel set of a CBF which is control-invariant. Typically, this set is an inner approximation of the maximal control invariant set\footnote{Here, we drop the keyword robust to indicate a disturbance free version of maximal robust control invariant set.}. 
We can visualize two different safe sets characterized by these methods as shown in Fig.~\ref{fig:geometric_connection}. The pink region contains all initial conditions that will inevitably (potentially when disturbance exists) enter failure states $\calF$ (red region) given bounded control signal.  This set is precisely captured by infinite-time BRT\footnote{Disturbance set $\calD = \emptyset$}. The complement of it (green region) is the maximal control invariant set which contains safe set $\calC$ characterized by a superlevel set of a CBF $b(\bfx)$. 

\textbf{Construct CBF from HJ-RA.} In general, constructing a CBF is challenging especially when control space is constrained. If the system dimension is low, \cite{CBF_magnus2018CA_CBF} suggests that we might construct CBF from HJ reachability analysis. Suppose there exist a 
reference controller $\bfu_{ref}(\bfx)$ is continuous differentiable can drive the agent toward goal configuration. However, this controller does not necessarily guarantee safety behavior. We can choose BRAT-based least-restrictive control policy ~\eqref{eq:safe_ctrl_brt} to be the evading maneuver controller. Then, we can define a candidate ZCBF, parameterized by $\bfu^{*}(\cdot)$ and performance function $l(\cdot)$ as follows:
\begin{equation}
	b\prl{\bfx; \pi_{BRT}, l } = \inf_{t \geq 0} \; l(\bfxi^{\pi_{BRT}}_{\bfx, 0}(t)) = V_{BRT}(\bfx)
\end{equation}
\begin{remark}
	The unsafe target sub-level set encoded by $\calF = \crl{l(\bfx) \leq 0}$ has to be convex and compact. The smoothness of this ZCBF candidate requires further investigation to be verified as an strict ZCBF.
\end{remark}

\textbf{Stability enforcement.} 
Among all properties of dynamical system, stability is probably the most important and it is the most crucial thing for developing controller and cornerstone of safety. In this paragraph, we want to discuss how stability related to those two safety methods. Let us recall some basics of Lyapunov stability theory. 
Suppose system ~\eqref{eq:sys_pf} is stabilized by $\bar{\bfu}(\bfx)$ and there is not disturbance $\calD = \emptyset$, then the system becomes autonomous 
\begin{equation}  \label{eq:sys_autonomous}
\dot{\bfx} = \bff(\bfx, \bar{\bfu}(\bfx)) 
\end{equation}
Without lost of generality, we assume the equilibrium point is at origin. 
\begin{definition} \label{def:stability}
	The equilibrium point $\bfx = \bf0$ of \eqref{eq:sys_autonomous} is 
	\begin{itemize}
		\item stable if, for each $\epsilon > 0$, there is $\delta = \delta(\epsilon) > 0$ such that 
		\begin{equation}
		\norm{\bfx(\bf0)} \leq \delta \implies \norm{\bfx(t)} \leq \epsilon, \quad \forall t \geq 0
		\end{equation}
		\item unstable if it is not stable 
		\item asymptotically stable (A.S.) if it is stable and $\delta$ can be chosen such that
		\begin{equation}
		\norm{\bfx(\bf0)} \leq \delta \implies \lim_{t \rightarrow \infty} \bfx(t) = \bf0
		\end{equation}
	\end{itemize}
\end{definition}
\begin{figure}[h]
	\centering
	\includegraphics[width=\linewidth]{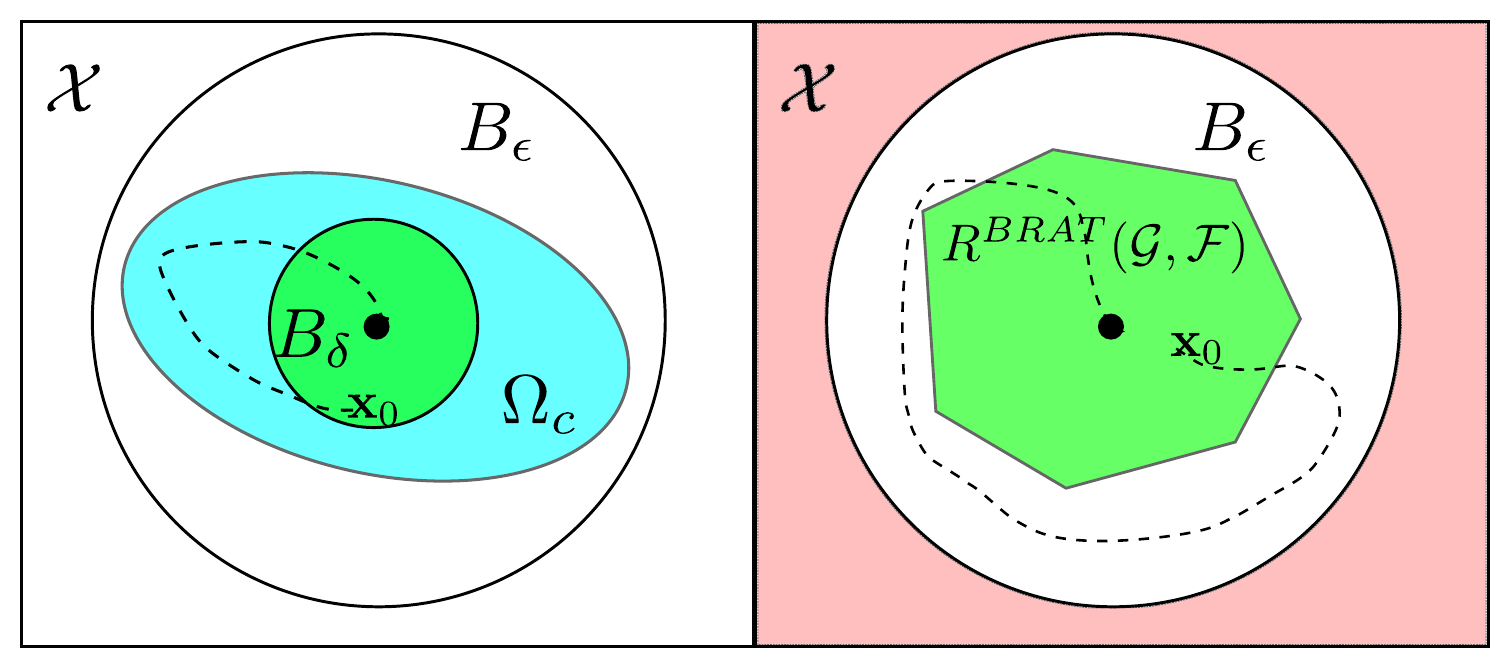}
	\caption{Geometric relationship of sets regarding stability. On the left, trajectory starting in the green ball $B_\delta$ will stay in invariant set $\Omega_c$ (blue ellipse) induced from Lyapunov function. Hence it will never leave out prescribed region in $B_\epsilon$ and finally will converge to equilibrium point. On the right, using HJ analysis can computes infinite-time BRAT (green polygon) contains all initial conditions that the agent will finally enter goal region $\calG = \crl{\bf0}$ without ever entering unsafe set (pink region) $\calF = B_\epsilon^c$. }
	\label{fig:stability_lyap_HJ}
\end{figure}
Using Lyapunov stability theory, one can show that if one can found a Lyapunov function $V(\bfx)$ for \eqref{eq:sys_autonomous}, then the equilibrium point at origin is A.S. In figure~\ref{fig:stability_lyap_HJ} is the geometric sets used in Proof of \cite[Theorem 4.1]{khalil2002nonlinear} on the left, where $B_\epsilon = \crl{\bfx :\norm{\bfx} \leq \epsilon}$, $B_\delta = \crl{\bfx :\norm{\bfx} \leq \delta}$ and $\Omega_c = \crl{\bfx: V(\bfx) \leq c}$ with $c = \min_{\norm{\bfx}=r} V(\bfx)$. From this theorem, we know that every trajectory starts inside the green circle will stays in invariant set $\Omega_c$ induced from Lyapunov function $V(\bfx)$ therefore by construction lies in $B_\epsilon$. Moreover, as time $t \rightarrow \infty$, the trajectory will convergence to equilibrium point. Next, we will show how CBF and HJ can be related to stability.

If the system dynamics is control-affine ~\eqref{eq:sys_ctrl_affine} and there exists upper bound over decay rate of Lyapunov function $V(\bfx)$, then Lyapunov function $V(\bfx)$ becomes a CLF and $\bar{u} \in \calK_{clf}$ is a particular feasible solution. Defining $b(\bfx) = c - V(\bfx)$, we get a CBF which is always feasible with  $\bar{u} \in \calK_{cbf}$. 
By setting goal set $G = \crl{\bf0}$ and unsafe set $\calF = B_\epsilon^c$, using HJ analysis, we can compute the infinite-time 
BRAT which captures all initial conditions that will eventually converge to equilibrium point without leaving specified bound 
$B_\epsilon$. In contrast to the region of attraction (ROA) estimation $B_\delta$ from Lyapunov analysis, $R^{BRAT}(\bf0, B_\epsilon^c)$ does not necessarily lies in some invariant set (blue ellipse for Lyapunov approach) as shown in Fig.~\ref{fig:stability_lyap_HJ}.  Moreover, the HJ analysis can give certificate beyond A.S. of the equilibrium point. Possible extensions can be:
\begin{itemize}
	\item checking finite-time horizon convergence property using finite-horizon BRAT
	\item certificating instability when $R^{BRAT}(\bf0, B_\epsilon^c)$ is empty
	\item ultimate boundedness \cite{khalil2002nonlinear} certification with and without disturbance 
	\item ROA estimation (maximal ROA, arbitrary shape ROA, ROA under input constraints, etc.)
\end{itemize}

%% file: tex/Comparison.tex
\section{Comparison}
We have already seen how CBF and HJ reachability analysis methods are connected to each other. In this section, we will focus on their difference in the following aspects: generality of system dynamics, difficulty of construction, computation cost. 
\begin{table*}[t]
	\centering
	\begin{tabular}{ |l|c|c| }
		\hline 
		& HJ-RA &  CBF \\ 
		\hline
		System dynamics 		& $\dot{\bfx} = \bff(t, \bfx, \bfu, \bfd)$	&  	 $\dot{\bfx} = \bff(\bfx) + \bfG(\bfx) \bfu$ \\ 
		Scalability	& poor (4 to 5 states) & good \\
		Realtimeness & slow (sec to hr) / offline & fast (ms) / online \\
		Constraints & time-varying set & state-dependent polytope \\
		Disturbance & bound per dimension & norm-based bound (RCBF) \\
		Optimality  & w.r.t target set & w.r.t. reference control / control effect \\
		\hline
	\end{tabular} 
	\caption{Comparison between Hamilton Jacobi reachability analysis and control barrier function based safe control methods in the scope of generality.} \label{tab:comp}
\end{table*}

\subsection{Generality} \label{sec:comp_gen}
The first question we want to find out is that when can we use CBF methods or HJ reachability analysis? This question seems to be quite simple at first glimpse. From Section~\ref{sec:background}, we know that Hamilton Jacobi reachability analysis applies to nonlinear system subject to bounded disturbance even for time-varying systems~\cite{HJ_time_varying}, whereas CBF approach can be used for deterministic control-affine system. But a closer look to each method will make this question much harder to answer. We  will consider generality in the following aspects: scalability, realtime-ness, disturbance handling, stability property and constraint enforcement.

\textbf{Scalability and Realtime-ness.}
Since HJ reachability solves PDE over grided state space using Dynamic Programming, a direct application of HJ reachability in most cases becomes intractable due to its exponentially scaled computational complexity. This limitation restricts its practical usage up to four/five states, however, recent work~\cite{HJA_decomposition_chen2017} exploits system structures to decompose the computation of reachable set into several small dimensional computations which significantly extends the usage of HJ reachability analysis. This technique has been successfully used in 10-D near hover quadrotors~\cite{HJ_tutorial,herbert2017fastrack}. In recent, researchers start using a neural network to compute BRS/BRT which could further improve its usage in real-time applications ~\cite{HJ_bansal2020deepreach}.
In contrast, Control barrier function method can handle high-order control-affine systems easily, for example on bipedal walking robots and multi-agent systems~\cite{CBF_ames2014rapidly,CBF_Ames2015_swarm}. Once a CBF is constructed, the safety constraints are converted to 
linear constraints in associated optimization problem which can be easily solved by off the shelf quadratic solver.

Due to the same reason for scalability, the computation burden for HJ reachability hinders its application in most robotic applications online. Therefore, offline pre-computation is often used~\cite{herbert2017fastrack}. To overcome this issue, warm-start and locally value function update techniques have been developed~\cite{HJ_bajcsy2019efficient}, which significantly improves computation efficiency and enables second-level reachable set computation in unknown environment for low-speed navigation task. In comparison, CBF method usually does not suffer from computation burden. Using efficient QP solver, the associated optimization problem \eqref{eq:CLF_CBF_QP} or \eqref{eq:MIX_CBF_QP} can be solved at hundreds of times per seconds. This makes it more favorable in time-sensitive robotic applications, such as Segway balancing, adaptive cruise control, walking robots~\cite{CBF_Segway,CBF_ames2017TAC,CBF_ames2014rapidly}. 

\textbf{Disturbance.}
It is paramount that controllers are robustly designed to ensure safety in the presence of disturbance. Such disturbance may come from uncertainty in modeling, imperfect sensory data as well external disturbance comes from environment. HJ reachability analysis treats control and disturbance in pursuer-evader differential games \cite{HJ_tutorial}, optimal control is obtained considering worst-case disturbance sequence might happens. Classical CBF formulation \eqref{def:CBF} does not implicitly consider the effect of disturbance and could potentially lead to unsafe behavior.  As CBF becomes a popular tool in many real applications, designing CBF against disturbance draws more and more attention. For example, robust CBF \cite{CBF_RCBF_jankovic2018robust} is designed considering worst-case disturbance
and \cite{CBF_MRCBF_dean2020guaranteeing} is aiming at designing robust CBF for measurement uncertainty. Researchers also consider using reinforcement-learning technique to quantity bounds when designing CBFs~\cite{CBF_choi2020_RL} to improve controller performance while guaranteeing safety.

\textbf{Constraints.}
In this paragraph, we want to discuss what kinds of constraints that CBF and HJ reachability analysis tool can deal with. 
From Section~\ref{sec:background}, we know that HJ handles constraints by defining unsafe or goal set. 
In the most general case, those sets can be time-varying characterized by Lipschitz continuous functions, e.g., signed distance functions \cite{HJ_time_varying}. 
From the formulation of CLF-CBF QP~\eqref{eq:CLF_CBF_QP}, CBF-based method can incorporate additional state-dependent linear constraints while enforcing safety using CBF.  In many robot applications, collision avoidance is one of the most important constraints. Although both methods require full knowledge of obstacle representations, using HJ analysis to deal with obstacle in working space is much easier than CBF construction. Suppose there exists $N$ obstacles in the environments, each of them is represented as $o_i  = \crl{g_i(\bfx) \geq 0}$. To compute BRT or BRAT, we can define unsafe set $\calF = \crl{\bfx : g(\bfx) \geq 0}$ with $g(\bfx) = \max_{i \in \brl{1,\ldots N}} g_i(\bfx)$ and result procedures are the same. For CBF based method this can be extremely challenging, researchers tries to using rational dynamics approximation and then using sum-of-square optimization tool to handle this problem \cite{CBF_barry2012CA}. Recent paper also tries to use SVM to learn control barrier function from sensor data \cite{CBF_SVM2020} however it requires a collision-free controller in advance. The summary of comparison for generality is shown in Table~\ref{tab:comp}.

\subsection{Difficulties and practical issues}
In this section, we will show the difficulties and practical issues about HJ reachability analysis and CBF based safety control methods. We will briefly talk about scalability and conservatism for HJ reachability analysis and then follow by practical issues of CBF approach such as construction difficulty and feasibility issue. At last, we will point out some common problems encountered in both methods.

\textbf{Limitations of HJ-RA.} Hamilton Jacobi reachability analysis is a great tool for computing reachable set and optimal control policy for complex system subject to adversarial disturbance. Scalability and conservatism are the two biggest problem hinders the usage of this method. As we discussed in Section~\ref{sec:comp_gen}, underlying computation complexity is scaled exponentially over state space dimension  and more details can be found in \cite{HJA_decomposition_chen2017,HJ_tutorial}. Another 
big issue is about conservatism built-in in the differential game setting. Disturbance bounds must be carefully chosen because the algorithm is only optimal when disturbance behaves in the worst case. Such assumption maybe fine for single agent operates in sparse environment and disturbance level is known to be relatively small. However, for two or multi-agent system considering all other agents to be adversarial is can be excessively conservative in practice.  Running cost from input is not considered in HJ-RA setup.

\textbf{Limitations of CBF based method.} The most important one might be how to construct such a function?. Unlike HJ analysis, one has to develop a valid CBF function which is challenging even for control-affine system without disturbance. For dynamical system that is not feedback-linearizable, there is not systematic way to construct a CBF~\cite{CBF_ames2019ECC}. The construction becomes even challenging when control signal $\bfu$ does not show up in the first derivative of barrier function $b(\bfx)$. Existing method like exponential CBF (ECBF) \cite{CBF_ECBF_nguyen2016ACC} assumes the system is feedback-linearizable and constant relative degree exist for all states, yet resulting design is initial condition dependent. Feasibility is another challenge lies in CBF based method. Safety linear constraints imposed by CBFs \eqref{eq:CBF_IE} are not guaranteed can be satisfied when multiple CBFs constraints exists. Only preliminary results are available for this question \cite{CBF_xu2018_CSBF} without considering disturbance and input constraints. The problem becomes worse when improper hyper-parameters are chosen.

\textbf{Common Issues.} Besides limitations mentioned above for each method, there  exist common issues as well. First of all, the most obvious one is both of them require accurate full knowledge of constraints representations. However, in many applications obstacles can only be partially observed at runtime using noisy measurements from onboard sensor. For multi-agent system, assuming full states are known might be impractical as well. Second, both HJ reachability and CBF methods often requires prefect knowledge of the model. Although one can price in model uncertainty by increasing disturbance magnitude, this may significantly reduce its performance. Another issue is about smoothness of resulting control signal, although preliminary results exists for CBF ~\cite{CBF_RCBF_jankovic2018robust}, the problem still considered to be challenging when input constraints are present.     

\begin{table*}[t]
	\centering
	\begin{tabular}{ |l|c|c| }
		\hline 
		& HJ-RA &  CBF \\ 
		\hline
		Scalability		& hard to scale	&  	 N/A \\ 
		Feasibility 	& N/A & no guarantee \\
		Design difficulty & low & high (especially for high relative degree CBF) \\
		Hyperparameter tuning & N/A & required \\
		Source of conservativeness & improper disturbance bound & CBF design, hyperparameters, norm-based disturbance bound \\
		Full knowledge of obstacle  & required & required \\
		\hline
	\end{tabular} 
	\caption{Practical issues comparison between Hamilton Jacobi reachability analysis and control barrier function based method.} \label{tab:issues}
\end{table*}

%% file: tex/CaseStudy.tex
\section{Case Study} \label{sec:case_study}
\begin{figure}[t!]
	\centering
	\includegraphics[width=0.49\linewidth]{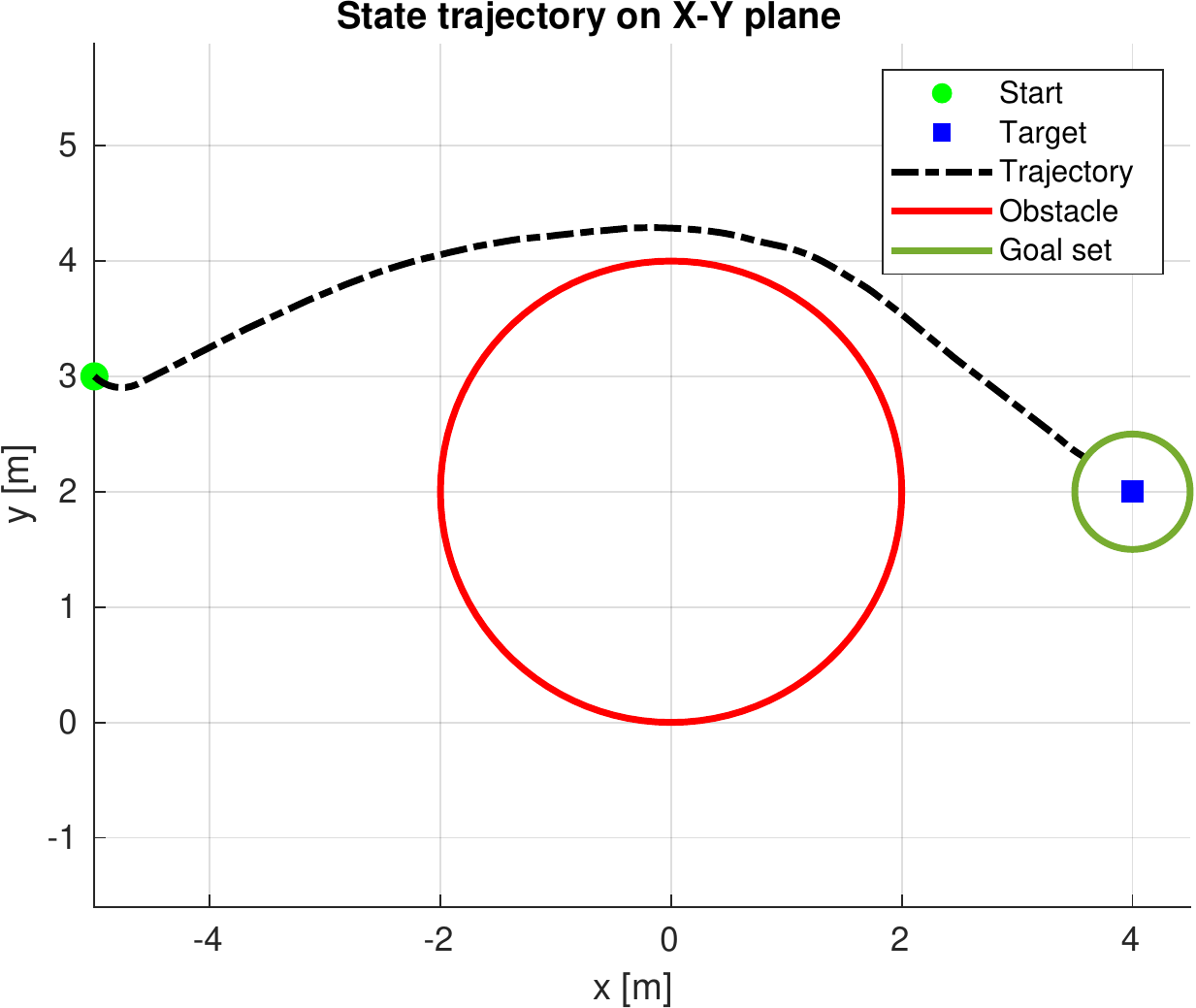}
	\includegraphics[width=0.49\linewidth]{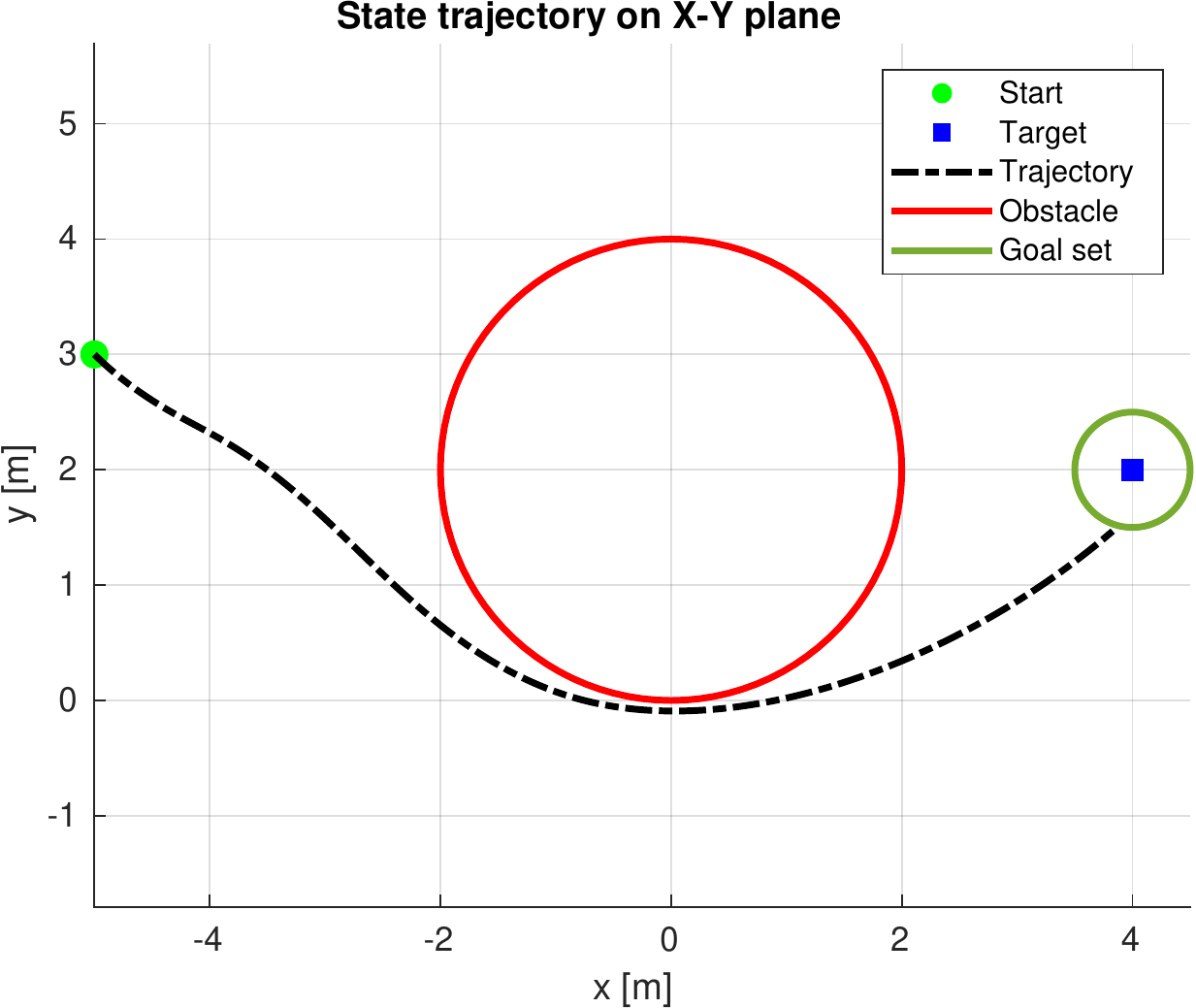}
	\caption{Simulation results of reach-avoid experiment for Dubins Car: HJ-RA (left) CBF (right). The trajectory projection of agents on working space is plotted as black dot line.	Both methods is able to drive the agent from start point (green dot) towards goal location (green circle) without entering obstacle (red circle). }
	\label{fig:sim_res}
\end{figure}
In this section, we will apply HJ analysis and CBF based method to a Dubins Car example in a reach-avoid problem setting. The robot is asked to travel from one point to another goal set $\calG$ while passing a obstacle $\calF$ in the middle of the way. The Dubins car model is defined as follows:
\begin{equation} \label{eq:sys_dubins_car}
\begin{bmatrix}
\dot{x} \\
\dot{y} \\
\dot{\theta}
\end{bmatrix} = \underbrace{\begin{bmatrix}
	v \cos \theta \\
	v \sin \theta \\
	0 
	\end{bmatrix}}_{\bff(\bfx)} + \underbrace{\begin{bmatrix}
	0\\
	0\\ 
	1
	\end{bmatrix}}_{\bfG} \underbrace{\begin{bmatrix}
	\omega 
	\end{bmatrix}}_{\bfu} + \underbrace{\begin{bmatrix}
	d_x \\ 
	d_y \\ 
	0
	\end{bmatrix}}_{\bfd},
\end{equation}
where $\bfx = \brl{x, y, \theta}^T$ with $\bfp_\bfx = (x, y)$ represents its position, input $\bfu = \brl{\omega}$ where $w \in \brl{-3, 3}$ denotes angular velocity with linear speed being constant at $v = 1$.

The experiment setup is as follows. The robot is started with initial condition $\bfx_0 = \brl{-5, 3, 0}^T$. The goal region is a ball center at $(4,2)$ with radius $0.5$. We place a circular obstacle centered at $(0,2)$ with radius $2$ between the start and goal region. The simulation results are shown in Fig.~\ref{fig:sim_res}, both methods are able to achieve the goal without collision. The result of HJ-RA is computed using  \href{https://github.com/HJReachability/helperOC}{helpOC toolbox} and tutorial can be found at \cite{HJ_tutorial}. For CBF-base method, we are using \href{https://github.com/HybridRobotics/CBF-CLF-Helper}{CBF-CLF-Helper}\footnote{The CLF and CBF in this repository is still experimental.}.  Note that, we do not impose any disturbance since classical CBF approach does not deal with uncertainty.

%% file: tex/Conclusion.tex
\section{Conclusion} 
\label{sec:conclusion} 
This report compares two popular safe control design methods: Hamilton Jacobi reachability analysis and control barrier function-based design.  The connection between HJ-RA and CBF is discussed in terms of safe set characterization, CBF construction using HJ-RA and stability enforcement. Comparison between HJ and classical CBF methods have been address in aspects of generality and practical issues. A case study on a low-order dynamical system is shown to demonstrate effectiveness of both methods. Through this discussion, we find out that HJ-RA has great flexibility and potential for low-order system. However, the scalability and demanding computation requirement hinder its usage. On the other hand, CBF based method is computation light and has much better scalability and have been used in many robot applications. The biggest problems of it are the difficulty of construction and feasibility concern.

%% file: root.bbl
\begin{thebibliography}{10}
\providecommand{\url}[1]{#1}
\csname url@rmstyle\endcsname
\providecommand{\newblock}{\relax}
\providecommand{\bibinfo}[2]{#2}
\providecommand\BIBentrySTDinterwordspacing{\spaceskip=0pt\relax}
\providecommand\BIBentryALTinterwordstretchfactor{4}
\providecommand\BIBentryALTinterwordspacing{\spaceskip=\fontdimen2\font plus
\BIBentryALTinterwordstretchfactor\fontdimen3\font minus
  \fontdimen4\font\relax}
\providecommand\BIBforeignlanguage[2]{{%
\expandafter\ifx\csname l@#1\endcsname\relax
\typeout{** WARNING: IEEEtran.bst: No hyphenation pattern has been}%
\typeout{** loaded for the language `#1'. Using the pattern for}%
\typeout{** the default language instead.}%
\else
\language=\csname l@#1\endcsname
\fi
#2}}

\bibitem{mitchell2002_thesis}
I.~M. Mitchell, ``Application of level set methods to control and reachability
  problems in continuous and hybrid systems.'' Ph.D. dissertation, Stanford
  University, 2002.

\bibitem{mitchell2007toolbox}
------, ``A toolbox of level set methods,'' \emph{UBC Department of Computer
  Science Technical Report}, 2007.

\bibitem{HJ_tutorial}
S.~Bansal, M.~Chen, S.~Herbert, and C.~J. Tomlin, ``Hamilton-jacobi
  reachability: A brief overview and recent advances,'' in \emph{IEEE
  Conference on Decision and Control (CDC)}, 2017, pp. 2242--2253.

\bibitem{HJA_decomposition_chen2017}
M.~Chen, S.~Herbert, and C.~J. Tomlin, ``{Exact and efficient Hamilton-Jacobi
  guaranteed safety analysis via system decomposition},'' in \emph{IEEE
  International Conference on Robotics and Automation (ICRA)}, 2017, pp.
  87--92.

\bibitem{herbert2017fastrack}
S.~L. Herbert, M.~Chen, S.~Han, S.~Bansal, J.~F. Fisac, and C.~J. Tomlin,
  ``{FaSTrack: A modular framework for fast and guaranteed safe motion
  planning},'' in \emph{IEEE Conference on Decision and Control (CDC)}, 2017,
  pp. 1517--1522.

\bibitem{CBF_ames2014control}
A.~D. Ames, J.~W. Grizzle, and P.~Tabuada, ``{Control barrier function based
  quadratic programs with application to adaptive cruise control},'' in
  \emph{IEEE Conference on Decision and Control (CDC)}, 2014, pp. 6271--6278.

\bibitem{CBF_ames2017TAC}
A.~D. Ames, X.~Xu, J.~W. Grizzle, and P.~Tabuada, ``{Control barrier function
  based quadratic programs for safety critical systems},'' \emph{IEEE
  Transactions on Automatic Control (TAC)}, vol.~62, no.~8, pp. 3861--3876,
  2017.

\bibitem{CBF_Ames2015_swarm}
U.~Borrmann, L.~Wang, A.~D. Ames, and M.~Egerstedt, ``{Control barrier
  certificates for safe swarm behavior},'' \emph{IFAC-PapersOnLine}, pp.
  68--73, 2015.

\bibitem{CBF_ames2014rapidly}
A.~D. Ames, K.~Galloway, K.~Sreenath, and J.~W. Grizzle, ``{Rapidly
  exponentially stabilizing control lyapunov functions and hybrid zero
  dynamics},'' \emph{IEEE Transactions on Automatic Control (TAC)}, vol.~59,
  no.~4, pp. 876--891, 2014.

\bibitem{CBF_quadrotor}
G.~Wu and K.~Sreenath, ``{Safety-critical control of a planar quadrotor},'' in
  \emph{American Control Conference (ACC)}, 2016, pp. 2252--2258.

\bibitem{CBF_ECBF_nguyen2016ACC}
Q.~Nguyen and K.~Sreenath, ``{Exponential control barrier functions for
  enforcing high relative-degree safety-critical constraints},'' in \emph{IEEE
  American Control Conference (ACC)}, 2016, pp. 322--328.

\bibitem{CBF_ames2019ECC}
A.~D. Ames, S.~Coogan, M.~Egerstedt, G.~Notomista, K.~Sreenath, and P.~Tabuada,
  ``{Control barrier functions: Theory and applications},'' in \emph{IEEE
  European Control Conference (ECC)}, 2019, pp. 3420--3431.

\bibitem{HJ_time_varying}
J.~F. Fisac, M.~Chen, C.~J. Tomlin, and S.~S. Sastry, ``Reach-avoid problems
  with time-varying dynamics, targets and constraints,'' in \emph{International
  conference on hybrid systems: computation and control (HSCC)}, 2015, pp.
  11--20.

\bibitem{CBF_Segway}
T.~Gurriet, A.~Singletary, J.~Reher, L.~Ciarletta, E.~Feron, and A.~Ames,
  ``{Towards a framework for realizable safety critical control through active
  set invariance},'' in \emph{International Conference on Cyber-Physical
  Systems (ICCPS)}.\hskip 1em plus 0.5em minus 0.4em\relax IEEE, 2018, pp.
  98--106.

\bibitem{CBF_magnus2018CA_CBF}
E.~Squires, P.~Pierpaoli, and M.~Egerstedt, ``{Constructive barrier
  certificates with applications to fixed-wing aircraft collision avoidance},''
  in \emph{IEEE Conference on Control Technology and Applications (CCTA)},
  2018, pp. 1656--1661.

\bibitem{khalil2002nonlinear}
H.~Khalil, \emph{{Nonlinear systems}}.\hskip 1em plus 0.5em minus 0.4em\relax
  Prentice Hall, 2002.

\bibitem{HJ_bansal2020deepreach}
S.~Bansal and C.~Tomlin, ``Deepreach: A deep learning approach to
  high-dimensional reachability,'' \emph{arXiv preprint arXiv:2011.02082},
  2020.

\bibitem{HJ_bajcsy2019efficient}
A.~Bajcsy, S.~Bansal, E.~Bronstein, V.~Tolani, and C.~J. Tomlin, ``{An
  efficient reachability-based framework for provably safe autonomous
  navigation in unknown environments},'' in \emph{IEEE Conference on Decision
  and Control (CDC)}, 2019, pp. 1758--1765.

\bibitem{CBF_RCBF_jankovic2018robust}
M.~Jankovic, ``Robust control barrier functions for constrained stabilization
  of nonlinear systems,'' \emph{Automatica}, vol.~96, pp. 359--367, 2018.

\bibitem{CBF_MRCBF_dean2020guaranteeing}
S.~Dean, A.~J. Taylor, R.~K. Cosner, B.~Recht, and A.~D. Ames, ``Guaranteeing
  safety of learned perception modules via measurement-robust control barrier
  functions,'' \emph{arXiv preprint arXiv:2010.16001}, 2020.

\bibitem{CBF_choi2020_RL}
J.~Choi, F.~Castaneda, C.~J. Tomlin, and K.~Sreenath, ``Reinforcement learning
  for safety-critical control under model uncertainty, using control lyapunov
  functions and control barrier functions,'' \emph{arXiv preprint
  arXiv:2004.07584}, 2020.

\bibitem{CBF_barry2012CA}
A.~J. Barry, A.~Majumdar, and R.~Tedrake, ``{Safety verification of reactive
  controllers for UAV flight in cluttered environments using barrier
  certificates},'' in \emph{IEEE International Conference on Robotics and
  Automation (ICRA)}, 2012, pp. 484--490.

\bibitem{CBF_SVM2020}
M.~Srinivasan, A.~Dabholkar, S.~Coogan, and P.~Vela, ``{Synthesis of control
  barrier functions using a supervised machine learning approach},'' in
  \emph{IEEE/RSJ International Conference on Intelligent Robots and Systems
  (IROS)}, 2020.

\bibitem{CBF_xu2018_CSBF}
X.~Xu, ``{Constrained control of input--output linearizable systems using
  control sharing barrier functions},'' \emph{Automatica}, vol.~87, pp.
  195--201, 2018.

\end{thebibliography}
